\documentclass[twocolumn, secnumarabic, amssymb, nobibnotes, aps, superscriptaddress]{revtex4-1}
\usepackage{graphicx}
\usepackage{multirow}
\usepackage[T1]{fontenc}

\begin{document}

\title{(Ga,Mn)As under pressure: a first-principles investigation}
\author{N. Gonzalez Szwacki}
\email{gonz@fuw.edu.pl}
\affiliation{Institute of Theoretical Physics, Faculty of Physics, University of Warsaw, ul. Pasteura 5, PL-02-093 Warszawa, Poland}
\author{Jacek A. Majewski}
\affiliation{Institute of Theoretical Physics, Faculty of Physics, University of Warsaw, ul. Pasteura 5, PL-02-093 Warszawa, Poland}
\author{T. Dietl}
\affiliation{Institute of Theoretical Physics, Faculty of Physics, University of Warsaw, ul. Pasteura 5, PL-02-093 Warszawa, Poland}
\affiliation{Institute of Physics, Polish Academy of Sciences, al. Lotników 32/46, PL-02-668 Warszawa, Poland}
\affiliation{WPI-Advanced Institute for Materials Research (WPI-AIMR), Tohoku University, 2-1-1 Katahira, Aoba-ku, Sendai 980-8577, Japan}

\begin{abstract}
Electronic and magnetic properties of Ga$_{1-}$\textit{$_{x}$}Mn\textit{$_{x}$}As, obtained from first-principles calculations employing the hybrid HSE06 functional, are presented for \textit{x}~=~6.25\% and 12.5\% under pressures ranging from 0 to 15~GPa. In agreement with photoemission experiments at ambient pressure, we find for \textit{x}~=~6.25\% that non-hybridized Mn-3\textit{d} levels and Mn-induced states reside about 5 and 0.4~eV below the Fermi energy, respectively. For elevated pressures, the Mn-3\textit{d} levels, Mn-induced states, and the Fermi level shift towards higher energies, however, the position of the Mn-induced states relative to the Fermi energy remains constant due to hybridization of the Mn-3\textit{d} levels with the valence As-4\textit{p} orbitals. We also evaluate, employing Monte Carlo simulations, the Curie temperature ($\textit{T}_{{\rm C}}$). At zero pressure, we obtain $\textit{T}_{{\rm C}}$~=~181~K, whereas the pressure-induced changes in $\textit{T}_{{\rm C}}$ are d$\textit{T}_{{\rm C}}$/d\textit{p}~=~+4.3~K/GPa for \textit{x}~=~12.5\% and an estimated value of d$\textit{T}_{{\rm C}}$/d\textit{p}~$\approx$~+2.2~K/GPa for \textit{x}~=~6.25\% under pressures up to 6~GPa. The determined values of d$\textit{T}_{{\rm C}}$/d\textit{p} compare favorably with d$\textit{T}_{{\rm C}}$/d\textit{p}~=~+(2--3)~K/GPa at \textit{p}~$\leq$~1.2~GPa found experimentally and estimated within the \textit{p}-\textit{d} Zener model for Ga$_{0.93}$Mn$_{0.07}$As in the regime where hole localization effects are of minor importance [M. Gryglas-Borysiewicz \textit{et al}., Phys. Rev. B \textbf{82}, 153204 (2010)]. 
\end{abstract}
\maketitle 

\section{INTRODUCTION}
The application of hydrostatic pressure serves as a unique tool for tuning the various essential parameters in semiconductors \cite{1}. High-pressure experiments and theoretical studies were important, for instance, in the delineation of the band structure of key semiconductors like Si, Ge, and the III-V compounds \cite{2}. This approach was also broadly used (both in theory and experiment) in the studies of bistability of donors (\textit{DX} centers) in GaAs and GaN \cite{3}.

For spintronic semiconductors, high-pressure measurements can play an important role in understanding the mechanisms that are responsible for the coupling between magnetic ions in dilute magnetic semiconductors (DMSs): the changes in volume of the solid influence the local exchange interaction between magnetic dopants providing information about the mechanism responsible for the observed macroscopic magnetization \cite{4}. Pressure can also help to elucidate the relationship between the coupling mechanism between magnetic dopants and the underlying band structure of the DMS. The states introduced by the dopants may have characteristics of the band edge states and follow the band edge directly when the pressure is applied or may tend to be highly localized, therefore, influenced by the entire Brillouin zone (the pressure dependence of deep levels does not follow any particular band edge) \cite{5}. Bearing in mind the sensitivity of the exchange interaction between magnetic dopants to interatomic distances in DMSs, their experimental and \textit{ab initio} theoretical studies under applied pressures are the most direct way to test different theoretical models that have been proposed to explain, for instance, the ferromagnetism in (Ga,Mn)As \cite{6}.

There exists a number of experimental studies on the variation of $\textit{T}_{{\rm C}}$ under pressure for DMSs: (Pb,Sn,Mn)Te \cite{7}, (In,Mn)Sb \cite{4,8}, (Sb,V)$_{2}$Te$_{3}$ \cite{9}, (Ga,Mn)As \cite{8,10}, and (Ge,Mn)Te \cite{11}. The results of those studies reveal that for samples with high concentrations of band holes, $\textit{T}_{{\rm C}}$ increases with an increasing hydrostatic pressure \textit{p} according to the expectations of the \textit{p}-\textit{d} Zener model \cite{12}. However, there are two worthwhile exceptions, i.e., the systems showing d$\textit{T}_{{\rm C}}$/d\textit{p}~$<$~0. The first is a narrow-gap topological insulator (Sb,V)$_{2}$Te$_{3}$ \cite{9}, in which rather than intraband excitations (i.e., the standard Ruderman-Kittel-Kasuya-Yosida mechanism), interband excitations (i.e., the Bloembergen-Rowland interactions) give a dominant contribution to the coupling between localized spins of transition metals (TMs) \cite{6}. The second is (Ga,Mn)As in the vicinity of the metal-insulator transition, where the increase in the \textit{p}-\textit{d} hybridization enhances hole localization and, hence, diminishes $\textit{T}_{{\rm C}}$ \cite{10}.

Not much has been accomplished from the \textit{ab initio} side so far. The pressure dependence of $\textit{T}_{{\rm C}}$ for Ga$_{0.95}$Mn$_{0.05}$As was studied by Bergqvist \textit{et} \textit{al}. \cite{13} within the local-density approximation (LDA) and the LDA+\textit{U} approach for pressures slightly crossing 7~GPa for which the direct-to-indirect band gap transition occurs for GaAs within the LDA \cite{14}. It was established (employing the LDA+\textit{U}) that $\textit{T}_{{\rm C}}$ increases for the decreasing lattice constant of Ga$_{0.95}$Mn$_{0.05}$As. This behavior was more pronounced within the mean-field approximation (MFA) than for the Monte Carlo (MC) simulations.

Here, we present first principles calculations for the structural, electronic, and magnetic properties of Mn\textit{$_{x}$}Ga$_{1-}$\textit{$_{x}$}As with \textit{x}~=~6.25\% and 12.5\%, exposed to pressures from 0 to 15~GPa. Therefore, we cover the whole range of pressures for which the zinc-blende structure of GaAs is stable \cite{15}. Although the focus of this paper is to study the pressure dependence of $\textit{T}_{{\rm C}}$, some other issues are also discussed, such as the influence of pressure onto the electronic and magnetic properties of Ga$_{0.9375}$Mn$_{0.0625}$As, the energetics of the formation of Mn pairs in GaAs, and finally, the pressure dependence of the exchange coupling between Mn ions both at purely substitutional and mixed substitutional-interstitial sites.

The paper is organized as follows. In Sec.~II the computational approach is discussed. In Sec.~III the results are presented, beginning with a discussion in Sec. IIIA of the structural, magnetic, and electronic properties of the studied DMS under pressure. The results concerning the cluster formation and the pressure dependence of $\textit{T}_{{\rm C}}$ are presented in Secs. IIIB and IIIC, respectively. We end with some concluding remarks in Sec.~IV.

\section{COMPUTATIONAL DETAILS}

The first-principles spin-polarized calculations are performed using a plane-wave basis and Troullier-Martins norm-conserving pseudo-potentials \cite{16,17} as implemented in the Quantum-ESPRESSO package \cite{18}. We employ the hybrid HSE06 exchange-correlation functional \cite{19,20} that appears particularly suitable for systems containing highly correlated localized electrons on TM d shells and itinerant band carriers. The plane-wave cutoff is set to be 40~Ry and a $8\times 8\times 8$ \textbf{\textit{k}}-point mesh for the Brillouin zone sampling is used. The calculations are done using a body centered cubic lattice of supercells containing 32 atoms. Its primitive vectors are $\vec{a}_{1} =a\left(-1,1,1\right)$, $\vec{a}_{2} =a\left(1,-1,1\right)$, and $\vec{a}_{3} =a\left(1,1,-1\right)$, where \textit{a} is the edge of the conventional unit cell. By replacing one Ga ion with a Mn ion, we obtain a ferromagnetic (FM) DMS with the Mn concentration of 6.25\%. We also study a system with two Mn ions in the Ga sublattice (it corresponds to Mn concentration of 12.5\%). Finally, for completeness, we study a system with one substitutional and one interstitial Mn ion in a chosen supercell. The calculated equilibrium values of lattice constant, bulk modulus, and pressure derivative of the bulk modulus for GaAs are $a=5.70\text{ \AA}$, $B=69.5{\rm \; GPa}$, and $B'=4$, respectively, and agrees well with earlier theoretical and experimental reports \cite{21,22}. Moreover, the computed GaAs lattice constant is close to the experimental values, 5.68~\AA\ (5.70~\AA) for Ga$_{1-}$\textit{$_{x}$}Mn\textit{$_{x}$}As with \textit{x}~$\approx$~6\% (\textit{x}~$\approx$~12\%) \cite{23}. The Murnaghan equation of state is used to estimate the theoretical pressures. Since HSE06 calculations are computationally very demanding in comparison to PBE ones, the internal coordinates of atoms of GaAs doped with Mn were optimized at the PBE level of theory.

Structural optimization is a prerequisite to a successful computation of the electronic structure. This is because relaxation around the Mn impurity could, in principle, influence the electronic structure substantially. We have explicitly allowed for structural relaxation by displacing each atom from its equilibrium position for unit cells at zero and higher pressures. We find, for instance, that at zero pressure, the introduction of one Mn ion has a large effect on the position of the first shell of As neighbors, where the Mn--As bond length is elongated by 3.5\% with respect to the Ga--As bond length of 2.469~\AA\ in GaAs. The tetrahedral symmetry is preserved at all pressures.

The MC simulations are performed using the VAMPIRE software package \cite{24}. The values of $\textit{T}_{{\rm C}}$ are calculated from the thermal average of magnetization vs. temperature curve. For these calculations, the cell size is chosen to be $22\times 22\times 22$ and for each temperature, we perform 240000 equilibration steps and the averaging loop is set to 240000 steps, which gives reasonable averaging. 

\section{RESULTS AND DISCUSSION}
\subsection{Structural, magnetic, and electronic properties}

For Ga$_{1-}$\textit{$_{x}$}Mn\textit{$_{x}$}As with \textit{x}~=~6.25\% and zero external pressure, we find a total magnetic moment of 4.27 Bohr-magnetons (\textit{$\mu$}$_{{\rm B}}$) per super-cell. This value is reduced comparing to the magnetic moment of 5\textit{$\mu$}$_{{\rm B}}$ for the free Mn atom by spins of band holes which are aligned according to the antiferromagnetic \textit{p}-\textit{d} exchange interaction. We note that the orbital contribution to the magnetic moment, brought about by the spin-orbit interaction, is below 0.1$\mu_{\text{B}}$ per Mn ion in (Ga,Mn)As \cite{Wadley:2010_PRB,Sliwa:2014_PRB}. For larger pressures, the total magnetic moment\textit{ }does not change significantly, for instance, for 7 and 15~GPa, it assumes the values of 4.23 and 4.19\textit{$\mu$}$_{{\rm B}}$, respectively.

\begin{figure} [t]
\centering
\includegraphics[width=8.6 cm]{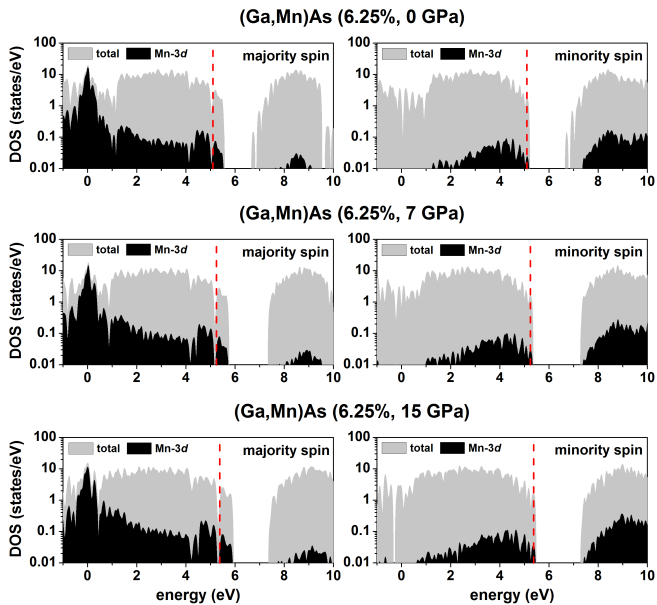}
\caption{(color online). Spin-resolved DOS in logarithmic scale of Mn$_{0.0625}$Ga$_{0.9375}$As (one Mn atom per supercell) at hydrostatic pressures of 0, 7, and 15~GPa. Light gray color denotes the total DOS whereas the Mn-3\textit{d} projected DOS is shown in black color. All the energy values are referred to the position of the main peak of the Mn-3\textit{d} states. The dashed vertical line indicates the Fermi level.}
\label{fig1}
\end{figure}

\begin{figure} [t]
\centering
\includegraphics[width=8.6 cm]{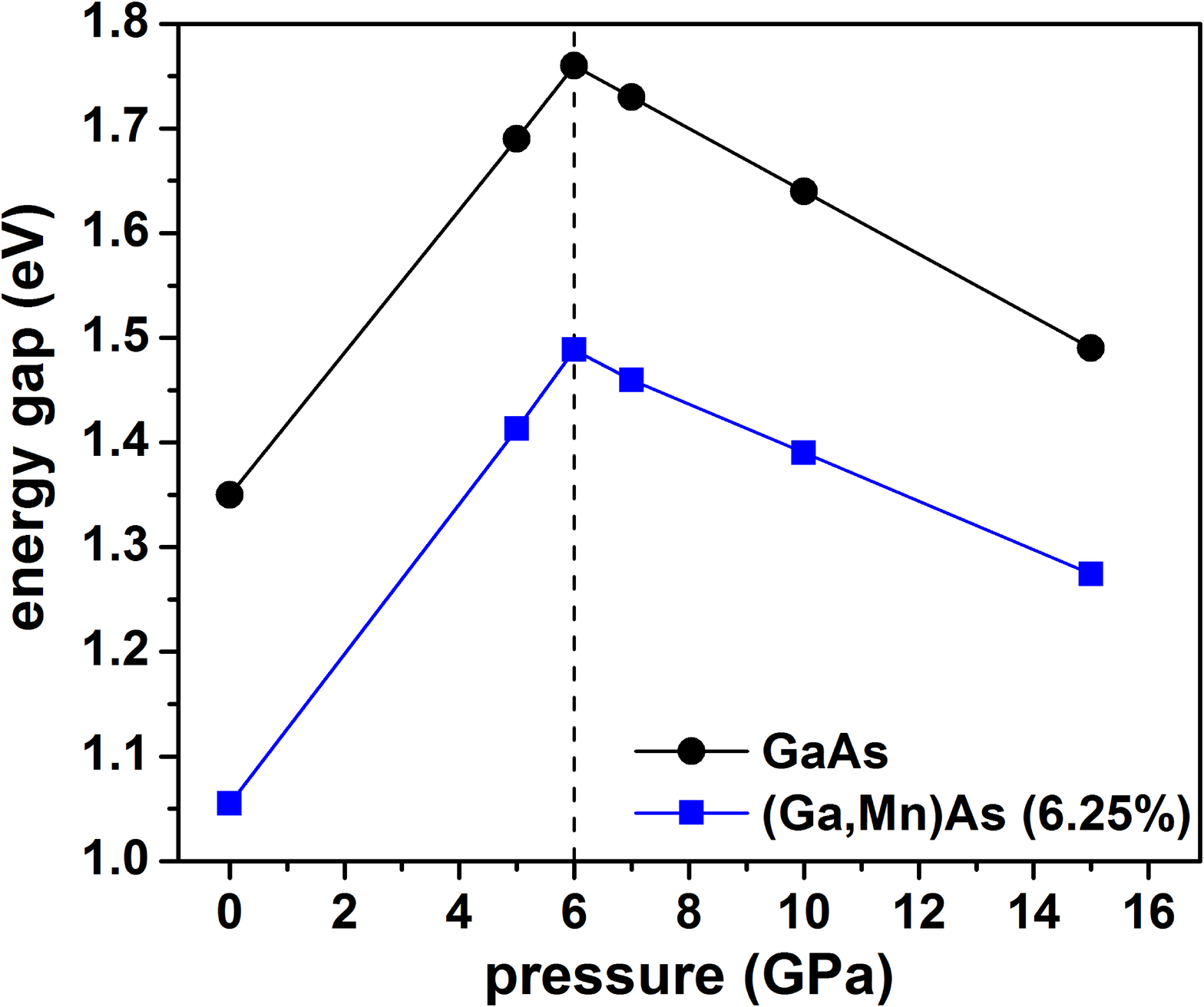}
\caption{(color online). Majority spin band gap of Ga$_{0.9375}$ Mn$_{0.0625}$As and the band gap of GaAs plotted as a function of pressure. In both cases the direct-to-indirect band gap transition occurs at $\sim$6~GPa.}
\label{fig2}
\end{figure}

Calculations of the density of states are presented for hydrostatic pressures of 0, 7, and 15~GPa in Fig.~\ref{fig1}. As seen, the majority and minority spin components show a band-gap, indicating that the introduction of the Mn ions does not destroy the semiconducting nature of the material. This is further shown in Fig.~\ref{fig2} where we plot the pressure dependence of the majority spin band gap compared with that of pure GaAs. In the isolated impurity limit, Mn in GaAs is a shallow acceptor, situated $\sim$0.1~eV above the valence band maximum \cite{25}. It hybridizes primarily with the valence band, and completely merges with it for a few percent of Mn (no impurity band). From Fig.~\ref{fig1}, it is apparent that the introduction of Mn to GaAs results in the Mn-induced spin-splitting of the valence band, where the valence band maximum of the majority spin is $\sim$0.3~eV above that of the minority spin. This prediction is in qualitative but not quantitative agreement with experiment \cite{26}, since experimentally a smaller splitting of $\sim$0.2~eV is expected for \textit{x}~$\approx$~6\% \cite{6}. It is, however, documented in the literature that the existing implementations of the density functional theory (DFT) tend to overestimate the \textit{p}-\textit{d} exchange splitting \cite{27}.

From Fig.~\ref{fig1}, we can also see that the Fermi level is not located entirely within the majority-band. This means that the system is not completely half metallic, and as a consequence, the total magnetic moment per Mn ion is larger than 4\textit{$\mu$}$_{{\rm B}}$. However, since our approach overestimates the exchange splitting, the real magnetic moment is closer to $5\mu_{\text{B}}$ than our theory predicts. The Mn-3\textit{d} states in Fig.~\ref{fig1} are for zero pressure located 5.1~eV below the Fermi level. This value is slightly overestimated with respect to recent experimental results in which the main peak of the Mn-3\textit{d} states was observed at 4.5~eV below the Fermi level \cite{28}. However, it should be noted that the theoretical results for the location of the Mn-3\textit{d} states strongly depend on the employed theoretical approach. The present results significantly improve previous DFT calculations that used the standard LDA or general gradient approximation (GGA) \cite{29}. In addition to the Mn-3\textit{d} states, we also find, in agreement with experiment \cite{28}, Mn-induced states centered about 0.4~eV below the Fermi energy. For elevated pressures (7 and 15~GPa in Fig.~\ref{fig1}), we obtain a pressure-induced shift of the Mn-3\textit{d} levels, Mn-induced states, and the Fermi energy towards higher energies, whereas the position of the Mn-induced states relative to the Fermi energy remains constant. This is an additional indication for hybridization of the Mn-3\textit{d} levels with the valence As-4\textit{p} orbitals. It is also clear from our computations that the Mn-derived spin-polarized feature in the majority spin band gap is not detached from the host valence band (cf. Refs. \cite{30,31}) and its pressure behavior is similar to that of the top of the valence band of the host material.

Finally, in Fig.~\ref{fig2} we plot the pressure dependence of the majority spin band gap for GaAs and Ga$_{0.9375}$Mn$_{0.0625}$As. GaAs has a direct band gap at ambient conditions. Under hydrostatic pressure, the upward shift of the conduction band minimum at the $\Gamma$ point eventually intersects the X minimum, and the material undergoes a direct-to-indirect band gap transition observed experimentally at $\sim$4~GPa \cite{32}. In Fig.~\ref{fig2}, we can see that the pressure behavior of the band gap of Ga$_{0.9375}$Mn$_{0.0625}$As is very similar to that of pure GaAs with a direct-to-indirect band gap transition at $\sim$6~GPa.

\subsection{Cluster formation}

We consider the case of the formation of clusters composed of two Mn ions in GaAs. It is well established \cite{33,34,35,36} that the magnetic ions may occupy both substitutional cation sites, ${\rm Mn}_{{\rm S}} $, and As- and Ga-coordinated tetrahedral interstitial sites, ${\rm Mn}_{{\rm I(As)}} $ and ${\rm Mn}_{{\rm I(Ga)}} $, respectively.

We start first with the configuration where the interstitial and substitutional ions are as far away from each other as possible within the supercell. For this case, we obtain that $({\rm Mn}_{{\rm S}} ,{\rm Mn}_{{\rm I(Ga)}} )$ is more stable than $({\rm Mn}_{{\rm S}} ,{\rm Mn}_{{\rm I(As)}} )$ by 0.11~eV. This result is congruent with previous computations \cite{33,34,35,36}.

Let us now focus on the formation of ${\rm Mn}_{{\rm S}}\text{--}{\rm Mn}_{{\rm S}}$ and ${\rm Mn}_{{\rm S}}\text{--}{\rm Mn}_{{\rm I}} $ complexes in GaAs. In both types of considered clusters, the Mn ions are the nearest neighbors (NNs). The dissociation energies for ${\rm Mn}_{{\rm S}}\text{--}{\rm Mn}_{{\rm S}} $, ${\rm Mn}_{{\rm S}}\text{--}{\rm Mn}_{{\rm I(As)}}$, and ${\rm Mn}_{{\rm S}}\text{--}{\rm Mn}_{{\rm I(Ga)}}$ pairs are defined as
\begin{equation}
\begin{array}{l}
{\Delta E_{{\rm SS}}=E({\rm Mn}_{{\rm S}},{\rm Mn}_{{\rm S}})-E({\rm Mn}_{{\rm S}}\text{--}{\rm Mn}_{{\rm S}})} \\
{\Delta E_{{\rm SI(As)}}=E({\rm Mn}_{{\rm S}},{\rm Mn}_{{\rm I(Ga)}})-E({\rm Mn}_{{\rm S}}\text{--}{\rm Mn}_{{\rm I(As)}})} \\
{\Delta E_{{\rm SI(Ga)}}=E({\rm Mn}_{{\rm S}},{\rm Mn}_{{\rm I(Ga)}})-E({\rm Mn}_{{\rm S}}\text{--}{\rm Mn}_{{\rm I(Ga)}})}
\end{array}
\end{equation}
respectively, where \textit{E} is the total energy of the 32- or 33-atom supercell. To calculate $E{\rm (Mn}_{{\rm S}} ,{\rm Mn}_{{\rm S}} {\rm )}$ and $E({\rm Mn}_{{\rm S}} ,{\rm Mn}_{{\rm I(Ga)}} )$, we place the magnetic ions as far away from each other as possible within the supercell. The largest possible distance between the magnetic ions is 5.701 and $6.179\text{ \AA}$ for ${\rm (Mn}_{{\rm S}} ,{\rm Mn}_{{\rm S}} {\rm )}$ and $({\rm Mn}_{{\rm S}} ,{\rm Mn}_{{\rm I(Ga)}})$, respectively. 
The calculated dissociation energies are 0.11, 0.18, and 0.42~eV for $\Delta E_{{\rm SS}} $, $\Delta E_{{\rm SI(As)}} $, and $\Delta E_{{\rm SI(Ga)}} $, respectively. Our results suggest that pairs involving ${\rm Mn}_{{\rm I}} $ are more likely to be present than purely substitutional clusters. It should be also noted that clusters involving Ga-coordinated interstitials are more stable than those involving As-coordinated interstitials. All mentioned values and some additional details for the Mn complexes in GaAs are summarized in Table~\ref{tab:t1}, where for comparison, we also include the isolated ${\rm Mn}_{2} $ dimer for which we predict an antiferromagnetic (AFM) coupling with a binding energy of 0.21~eV in agreement with more sophisticated calculations involving quantum-chemical methods \cite{37}. The activation energy for the out diffusion of charged ${\rm Mn}_{{\rm I}} $ ions that are bound in interstitial-substitutional complexes is a sum of the dissociation energy plus the energy barrier for diffusion of charged ${\rm Mn}_{{\rm I}} $ ions between interstitial sites \cite{34,36}. It should be, however, noted that the computed by us dissociation energies for clusters involving neutral ${\rm Mn}_{{\rm I}} $ ions are equal to activation energies because of the lack of the migration barrier for swapping of neutral ${\rm Mn}_{{\rm I}} $ ions between interstitial sites \cite{36}.

\begin{table} [t]
\centering
\caption{Dissociation energies, interionic distances, and total energy differences per ion between FM and AFM states for various configurations of Mn complexes in GaAs at zero pressure. Results for an isolated Mn dimer ${\rm (Mn}_{2} )$ are also included.}
\begin{ruledtabular}
\begin{tabular}{c|ccc}
& dissociation energy & \multirow{2}{*}{\textit{d} (\AA)} & $E_{{\rm FM}}-E_{{\rm AFM}}$ \\
& (eV) & & (meV/Mn) \\ [0.5ex]
\hline
${\rm Mn}_{{\rm S}}\text{--}{\rm Mn}_{{\rm I(Ga)}} $ & 0.42 & 2.699 & 69 \\ 
${\rm Mn}_{{\rm S}}\text{--}{\rm Mn}_{{\rm I(As)}} $ & 0.18 & 2.879 & 33 \\ 
Mn$_{2}$ & 0.21 & 3.243 & 30 \\ 
${\rm Mn}_{{\rm S}}\text{--}{\rm Mn}_{{\rm S}} $ & 0.11 & 3.994 & -39 \\
\end{tabular}
\end{ruledtabular}
\label{tab:t1}
\end{table}

\begin{figure} [t]
\centering
\includegraphics[width=8.6 cm]{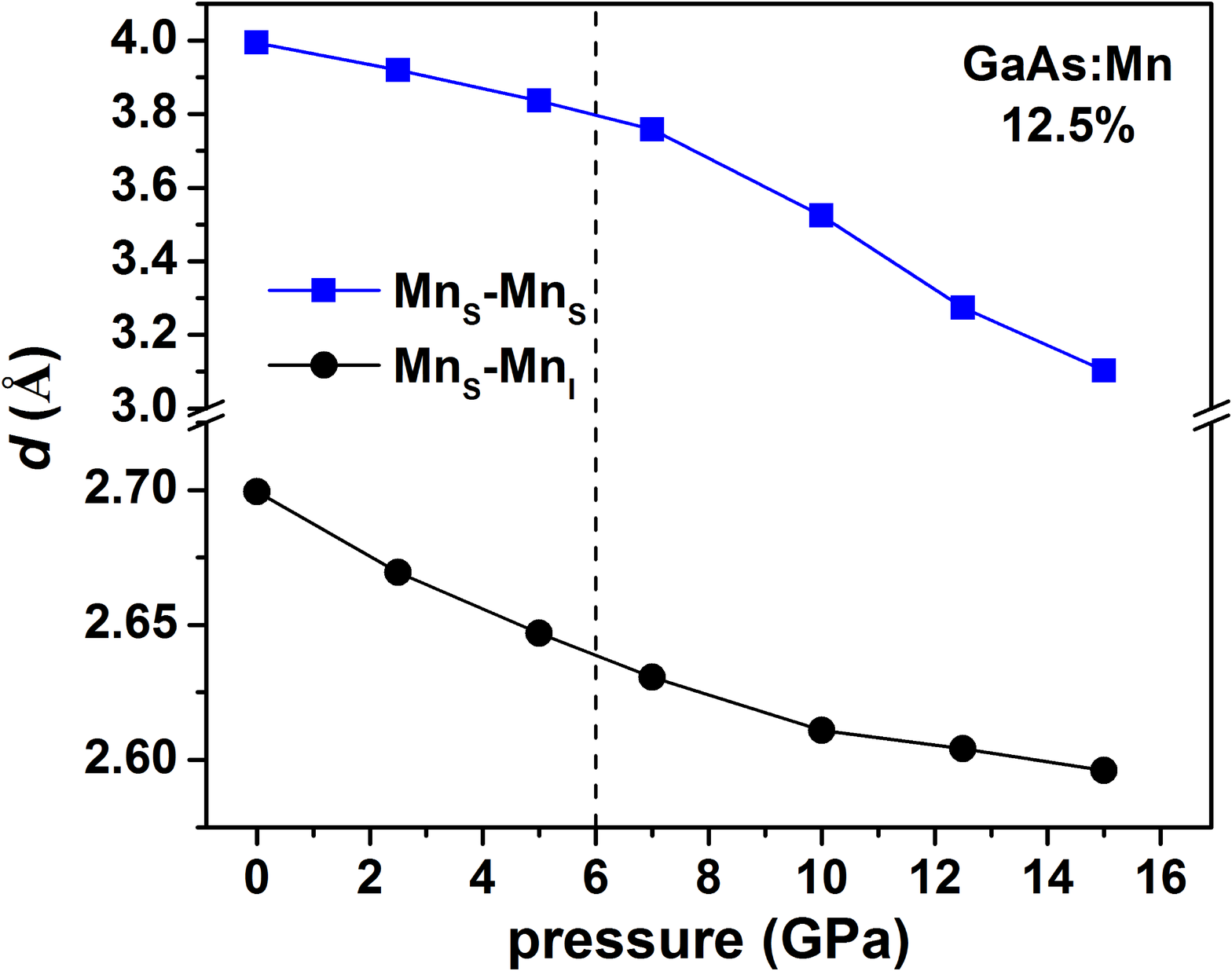}
\caption{(color online). Nearest neighbor distances \textit{d} plotted as a function of pressure for ${\rm Mn}_{{\rm S}}\text{--}{\rm Mn}_{{\rm I(Ga)}} $ (circles) and ${\rm Mn}_{{\rm S}}\text{--}{\rm Mn}_{{\rm S}} $ (squares) pairs in GaAs.}
\label{fig3}
\end{figure}

Finally, in Fig.~\ref{fig3}, we show the Mn--Mn NN distance \textit{d} as a function of pressure. It is clear from the figure that the variation with pressure of $d({\rm Mn}_{{\rm S}}\text{--}{\rm Mn}_{{\rm I(Ga)}})$ is much smaller than that of $d({\rm Mn}_{{\rm S}}\text{--}{\rm Mn}_{{\rm S}})$. For comparison, at zero pressure the unrelaxed values of $d({\rm Mn}_{{\rm S}}\text{--}{\rm Mn}_{{\rm S}})$ and $d({\rm Mn}_{{\rm S}}\text{--}{\rm Mn}_{{\rm I(Ga)}})$ --where atoms are placed into ideal lattice positions-- are 4.032 and 2.469~\AA, respectively (the relaxed values are listed in Table~\ref{tab:t1}).

\subsection{Curie temperature}

\begin{figure} [b]
\centering
\includegraphics[width=8.6 cm]{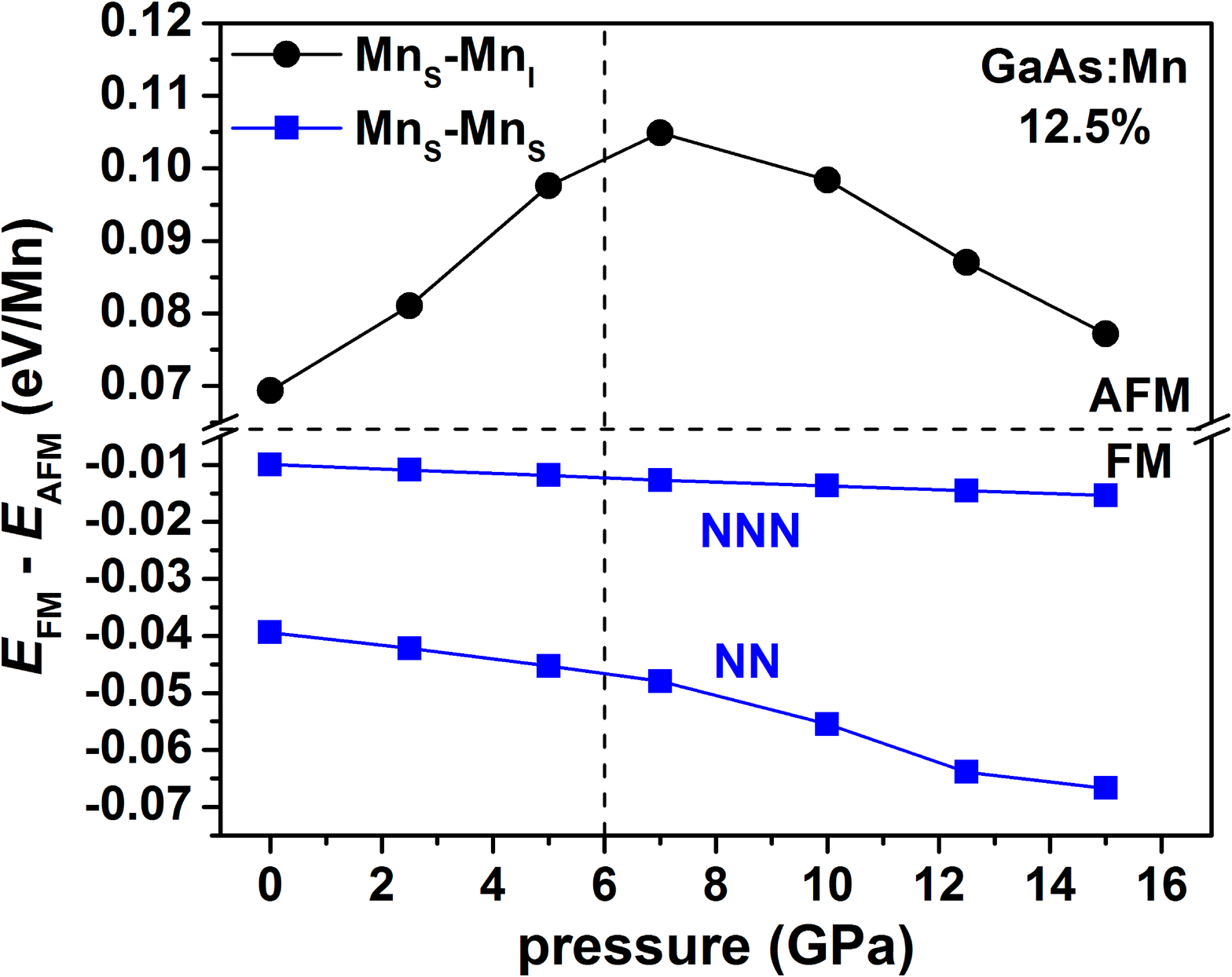}
\caption{(color online). Total energy difference per ion between the FM and AFM configurations for ${\rm Mn}_{{\rm S}}\text{--}{\rm Mn}_{{\rm I(Ga)}} $ (circles) and ${\rm Mn}_{{\rm S}}\text{--}{\rm Mn}_{{\rm S}} $ (squares) pairs in GaAs as a function of pressure. For ${\rm Mn}_{{\rm S}}\text{--}{\rm Mn}_{{\rm S}} $, we have considered NN and NNN positions in the Ga sublattice.}
\label{fig4}
\end{figure}

It is well established \cite{6} that one of the factors limiting high-temperature ferromagnetism is the presence of ${\rm Mn}_{{\rm S}}\text{--}{\rm Mn}_{{\rm I}}$ complexes in GaAs in which the Mn ions exhibit an AFM coupling. This coupling is not only preserved but tends to increase in strength with pressure (for pressures up to $\sim$7~GPa). This is shown in Fig.~\ref{fig4}, where we plotted the total energy difference, $E_{{\rm FM}} -E_{{\rm AFM}}$, between the FM and AFM configurations for ${\rm Mn}_{{\rm S}}\text{--}{\rm Mn}_{{\rm I(Ga)}}$ and ${\rm Mn}_{{\rm S}}\text{--}{\rm Mn}_{{\rm S}}$ pairs as a function of pressure. For the substitutional Mn ions, we have considered NN and next nearest neighbor (NNN) positions in the Ga sublattice.

\begin{figure} [t]
\centering
\includegraphics[width=8.6 cm]{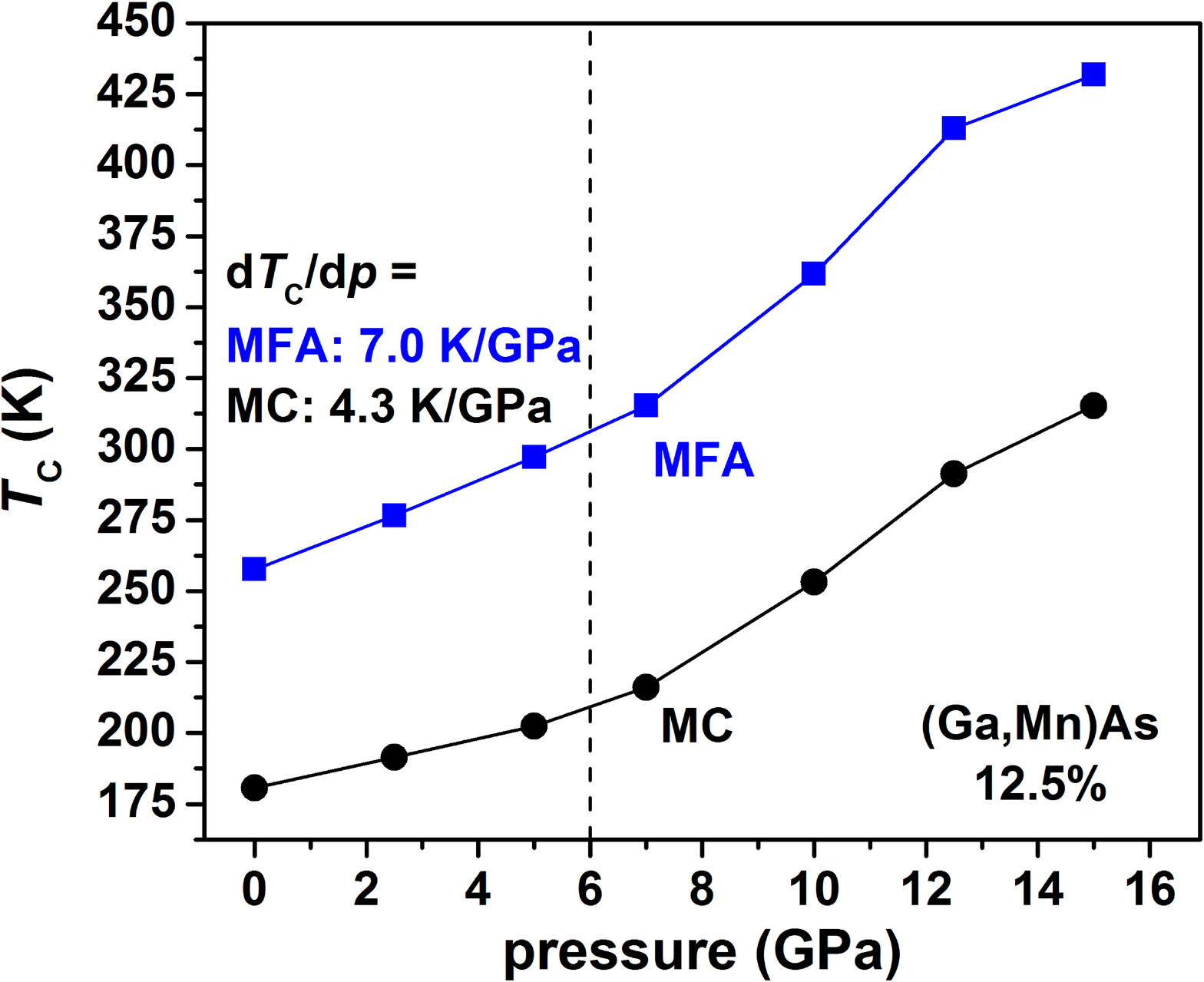}
\caption{(color online). Curie temperature of (Ga,Mn)As calculated within MFA (squares) and MC (circles) methods as a function of pressure. The given values of the pressure coefficients correspond to the region below 6~GPa marked by a dashed line.}
\label{fig5}
\end{figure}

The energy difference $E_{{\rm FM}}-E_{{\rm AFM}}$ for the substitutional ions can be used to evaluate the exchange interaction $J_{ij} $ between \textit{i} and \textit{j}-site local spins. The exchange energy for a system of interacting atomic moments is given by the effective classical Heisenberg Hamiltonian $H=-\sum _{i\ne j}J_{ij}\vec{S}_{i}\vec{S}_{j}=-2\sum_{i<j}J_{ij}\vec{S}_{i}\vec{S}_{j}$, where in our case $S_{i}=S_{j}=S={5\mathord{\left/{\vphantom {5 2}} \right. \kern-\nulldelimiterspace} 2} $. For ferromagnetically coupled Mn ions at NN substitutional positions $-2J_{{\rm NN}}S^{2} =E_{{\rm FM}} -E_{{\rm AFM}} $. For zero pressure $E_{{\rm FM}}-E_{{\rm AFM}}=-39\; {\rm meV/Mn}$, therefore we obtain that $J_{{\rm NN}} S^{2}=20{\rm \; meV}$; the calculated in a similar way $J_{{\rm NNN}} S^{2}$ for NNNs gives a modest value of 5~meV \cite{38}. The Curie temperature in the MFA is given by
\begin{equation}
\begin{array}{l}
{T_{{\rm C}} =\left({2\mathord{\left/ {\vphantom {2 3k_{{\rm B}} }} \right. \kern-\nulldelimiterspace} 3k_{{\rm B}} } \right)\cdot \left(x\sum _{j}z_{j} J_{0j} S^{2} \right)=} \\ 
{=\left({2S^{2} \mathord{\left/ {\vphantom {2S^{2} 3k_{{\rm B}} }} \right. \kern-\nulldelimiterspace} 3k_{{\rm B}} } \right)\cdot {\left(12J_{{\rm NN}} +6J_{{\rm NNN}} \right)\mathord{\left/ {\vphantom {\left(12J_{{\rm NN}} +6J_{{\rm NNN}} \right) 8}} \right. \kern-\nulldelimiterspace} 8} }
\end{array}
\end{equation}
where $x$ is the concentration of Mn and $z_{j}$ denotes the number of sites on the \textit{j}-th shell. The Curie temperature versus pressure is plotted in Fig.~\ref{fig5}. From the slope of the MFA curve, we can estimate the derivative of temperature with respect to pressure d$\textit{T}_{{\rm C}}$/d\textit{p}~=~7~K/GPa for pressures close to zero. A linear dependence of the MFA curve extends, however, up to $\sim$6~GPa. Assuming a linear dependence of the Curie temperature versus manganese concentration, we can estimate that for a $6.25\% $ concentration of Mn d$\textit{T}_{{\rm C}}$/d\textit{p}~= 3.5~K/GPa. Beyond 6~GPa the Curie temperature increases even faster, reaching the room temperature for $\sim$7~GPa.

We have also carried out MC simulations using the same exchange parameters $J_{{\rm NN}} $ and $J_{{\rm NNN}} $ as determined above. From Fig.~\ref{fig5}, we can see that the pressure behavior of $\textit{T}_{{\rm C}}$ calculated by means of MC is similar to that obtained from MFA calculations. However, the critical temperature is shifted down, for all pressures, by an average value of 26\%. For pressures close to 0~GPa, d$\textit{T}_{{\rm C}}$/d\textit{p}~=~4.3~K/GPa , and an estimated value for \textit{x}~=~6.25\% would be d$\textit{T}_{{\rm C}}$/d\textit{p}~=~2.2~K/GPa, which is in the range of values obtained experimentally d$\textit{T}_{{\rm C}}$/d\textit{p}~=~(2--3)~K/GPa at \textit{p}~$\leq$~1.2~GPa and estimated within the \textit{p}-\textit{d} Zener model for Ga$_{0.93}$Mn$_{0.07}$As \cite{10}.

\section{CONCLUSIONS}
In this paper, we have presented the pressure dependence of the Curie temperature for Ga$_{1-}$\textit{$_{x}$}Mn$_{x}$As with an experimentally realistic Mn contents of \textit{x}~=~6.25\% and \textit{x}~=~12.5\%, using the hybrid HSE06 functional for the description of exchange and correlation effects. In agreement with photoemission experiments, we have found for \textit{x}~=~6.25\% and ambient pressure that the non-hybridized Mn-3\textit{d} levels and Mn-induced states reside about 5 and 0.4~eV below the Fermi energy, respectively. For elevated pressures, the Mn-3\textit{d} levels, Mn-induced states, and the Fermi level shift towards higher energies, however, the position of the Mn-induced states relative to the Fermi energy remains constant due to hybridization of the Mn-3\textit{d} levels with the valence As-4\textit{p} orbitals. It follows from our computations that the Mn-derived spin-polarized feature in the majority spin band gap is not detached from the host valence band and its pressure behavior is similar to that of the top of the valence band of the host material. We have also found that $\textit{T}_{{\rm C}}$ at zero pressure is 181 and 258~K for MC and MFA calculations for \textit{x}~=~12.5\%, respectively, while increases linearly under pressures up to 6~GPa. The estimated for \textit{x}~=~6.25\% pressure-induced changes in $\textit{T}_{{\rm C}}$ are +2.2 and +3.5~K/GPa for MC and MFA calculations, respectively. The determined values of d$\textit{T}_{{\rm C}}$/d\textit{p} compare well with those found experimentally and estimated within the \textit{p}-\textit{d} Zener model for Ga$_{0.93}$Mn$_{0.07}$As in the regime where hole localization effects are of smaller importance \cite{10}.

\begin{acknowledgements}
The work was supported by the European Research Council through the FunDMS Advanced Grant within the ``Ideas'' Seventh Framework Programme of the EC and National Center of Science in Poland (Decision No. 2011/02/A/ST3/00125). We acknowledge the access to the computing facilities of the Interdisciplinary Center of Modeling at the University of Warsaw.
\end{acknowledgements}

\bibliographystyle{apsrev4-1}
\bibliography{v11}

\begin{thebibliography}{40}
\bibitem [{38()}]{38}%
  \BibitemOpen
  \href@noop {} {\ }\bibinfo {note} {In the 32-atom unit cell, each Mn ion has
  two NNN; therefore, in order to calculate the NNN exchange constat, we have
  divided $E_{{\rm FM}}-E_{{\rm AFM}}$ additionally by a factor of
  2.}\BibitemShut {Stop}%
\end{thebibliography}

\end{document}